\documentstyle[12pt,twoside,fleqn,espcrc1,epsfig]{article}

\newcommand{\AmS}{{\protect\the\textfont2
  A\kern-.1667em\lower.5ex\hbox{M}\kern-.125emS}}

\title{Hydrodynamic simulation of elliptic flow\thanks{This 
work was supported by BMBF, DFG and GSI.}}

\author{P.F.~Kolb\address{CERN/TH, CH-1211 Geneva 
        23}$^{\mbox{\scriptsize ,b}}$, J.~Sollfrank\address{Institut 
    f\"ur Theoretische Physik, Universit\"at Regensburg, D-93040 Regensburg}, 
    P.V.~Ruuskanen\address{Department of Physics, University of
      Jyv\"askyl\"a, FIN-40351 Jyv\"askyl\"a}, and
    U.~Heinz$^{\mbox{\scriptsize a}}$}

\begin{document}
% typeset front matter
\maketitle

\begin{abstract}
We use a hydrodynamic model to study the space-time evolution 
transverse to the beam direction in ultrarelativistic heavy-ion 
collisions with nonzero impact parameters. We focus on the 
influence of early pressure on the development of radial and 
elliptic flow. We show that at high energies elliptic flow is 
generated only during the initial stages of the expansion while 
radial flow continues to grow until freeze-out. Quantitative 
comparisons with SPS data from semiperipheral Pb+Pb collisions 
suggest the applicability of hydrodynamical concepts already 
$\approx$ 1 fm/c after impact.
\end{abstract}

%%%%%%%%%%%%%%%%%%%%%%%%%%%%%%%%%%%%%%%%%%%%%%%%%%%%%%%%%%%%%%%%%%%%%%%%
\section{Hydrodynamic model with longitudinal boost invariance}
%%%%%%%%%%%%%%%%%%%%%%%%%%%%%%%%%%%%%%%%%%%%%%%%%%%%%%%%%%%%%%%%%%%%%%%%

The transverse expansion dynamics in non-central heavy-ion collisions 
at SPS energies has recently attracted much attention 
\cite{Appe98,Agga98,Cere98,Olli92,Sorg99,TS99,Kolb99_2}. 
We here study it within the hydrodynamic model. In order to reduce
the complexity of the numerical task we follow \cite{Olli92}
and implement analytically Bjorken scaling flow with $v_z=z/t$ 
in the longitudinal direction and only solve the transverse
dynamics numerically. The Bjorken ansatz holds exactly at infinite
beam energy, but properly restricted to a finite rapidity interval it 
is phenomenologically successful also at SPS and AGS energies 
\cite{Dobl99}. It breaks down, however, near target and projectile 
rapidities; using it we can therefore reliably compute the 
transverse expansion only near midrapidity. 

The system of hydrodynamic equations is closed by an equation of state
(EOS) $p(e,n)$ giving the pressure as a function of energy and baryon 
density. Hydrodynamics thus provides a direct relation between the EOS
and the dynamical evolution of the system. To study the dynamical 
effects of a softening of the EOS in the neighborhood of a phase 
transition to quark-gluon plasma we use three different equations 
of state. EOS~I is the hard equation of an ideal ultrarelativistic 
gas, $p=e/3$. EOS~H is the much softer EOS for a gas of interacting 
hadron resonances; for $n\approx 0$ it satisfies $p\approx 0.15\,e\,$. 
A Maxwell construction between these two EOS, adding a bag pressure 
$B^{1/4}=230$ MeV, results in EOS~Q which has a phase transition at 
$T_{\rm cr}(n=0) = 164$ MeV with a latent heat of 1.15 GeV/fm$^3$ 
\cite{Soll97}. The system is frozen out at a fixed decoupling 
temperature $T_{\rm dec}$, and all unstable resonances are allowed 
to decay before we compare with experimental data.

%%%%%%%%%%%%%%%%%%%%%%%%%%%%%%%%%%%%%%%%%%%%%%%%%%%%%%%%%%%%%%%%%%%%%%%%
\section{Space time evolution of the reaction zone}
%%%%%%%%%%%%%%%%%%%%%%%%%%%%%%%%%%%%%%%%%%%%%%%%%%%%%%%%%%%%%%%%%%%%%%%%

We initialize the reaction zone with transverse energy and baryon 
density profiles which are taken to be proportional to the transverse
density of wounded nucleons calculated from the Glauber model
\cite{Olli92,Kolb99}. The initial configuration is thus parametrized
by the equilibration time $\tau_0$ and the maximum energy and baryon 
densities $e_0$ and $n_0$ in central  \\
%%%%%%%%%%%%%%%%%%%%%%%%% Fig. 1a-c %%%%%%%%%%%%%%%%%%%%%%%%%%%%%%%%%%%%%
\begin{minipage}[c]{6.3cm}
\hspace*{-.0cm}
\begin{minipage}{6.2cm}
  \epsfxsize 6.10cm
  \epsfbox{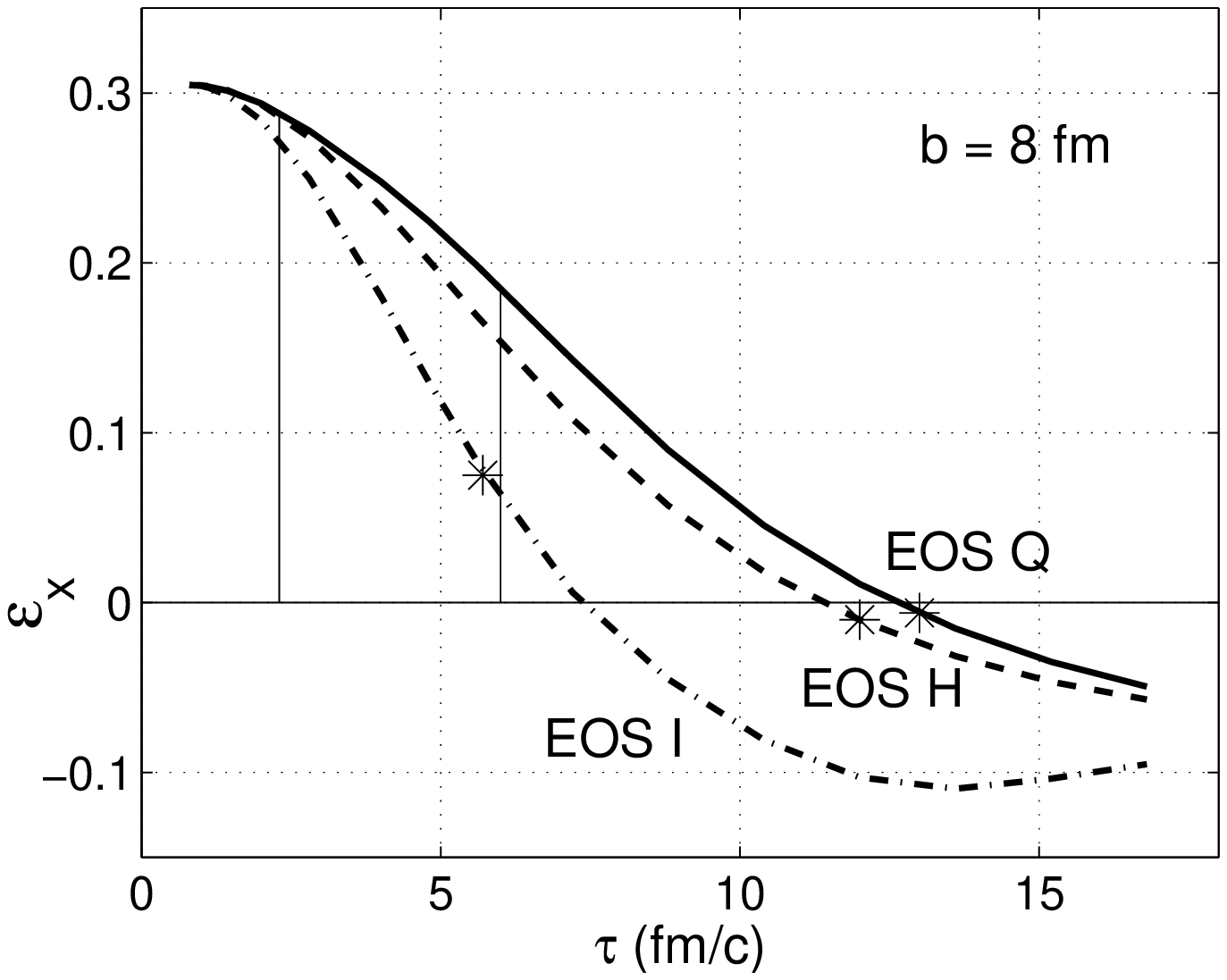}
\end{minipage}\\
\hspace*{-.8mm}
\begin{minipage}{62mm}
  \epsfxsize 61.5mm
  \epsfbox{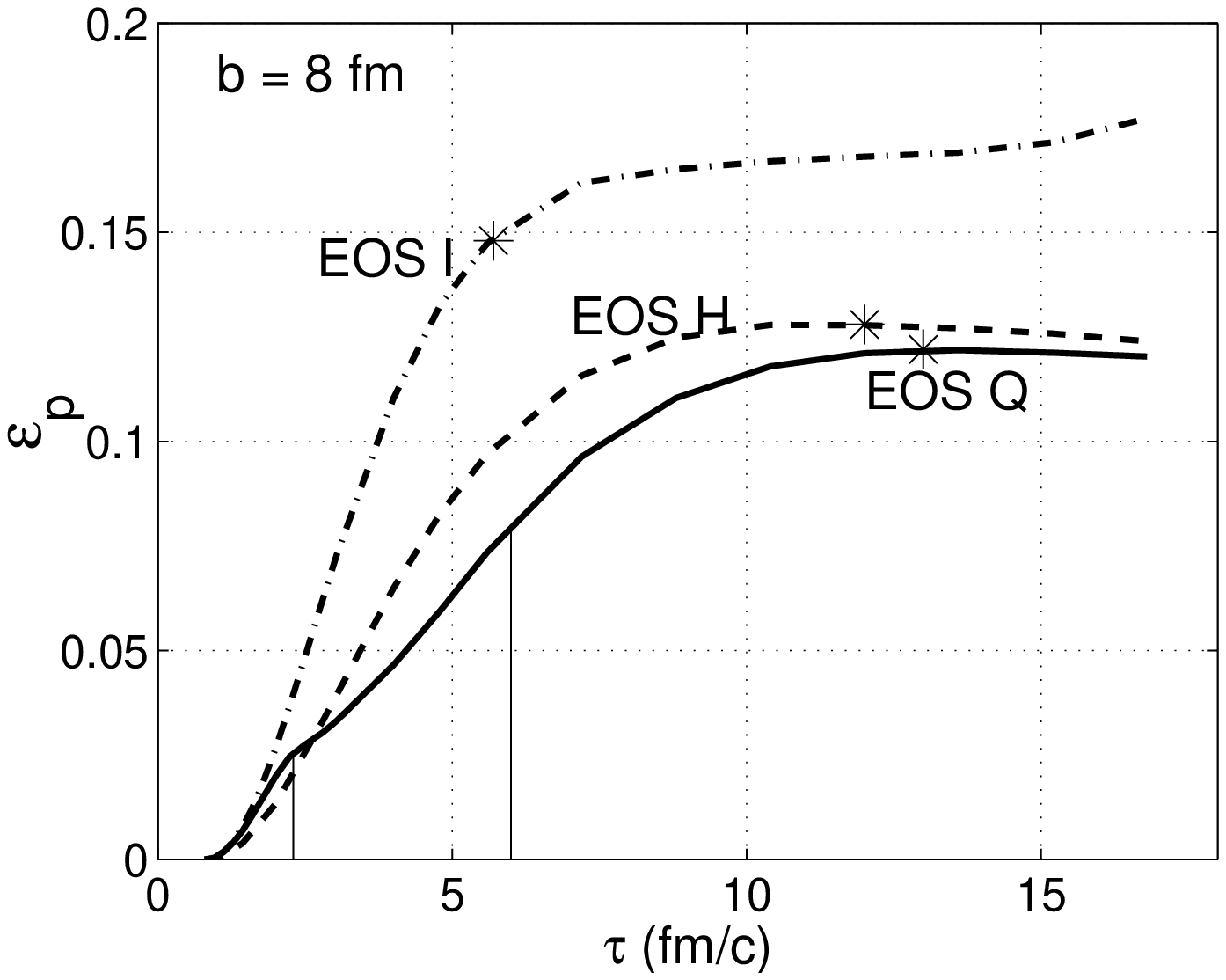}
\end{minipage}\\
\hspace*{0.16cm}
\begin{minipage}{62mm}
  \epsfxsize 59.7mm
  \epsfbox{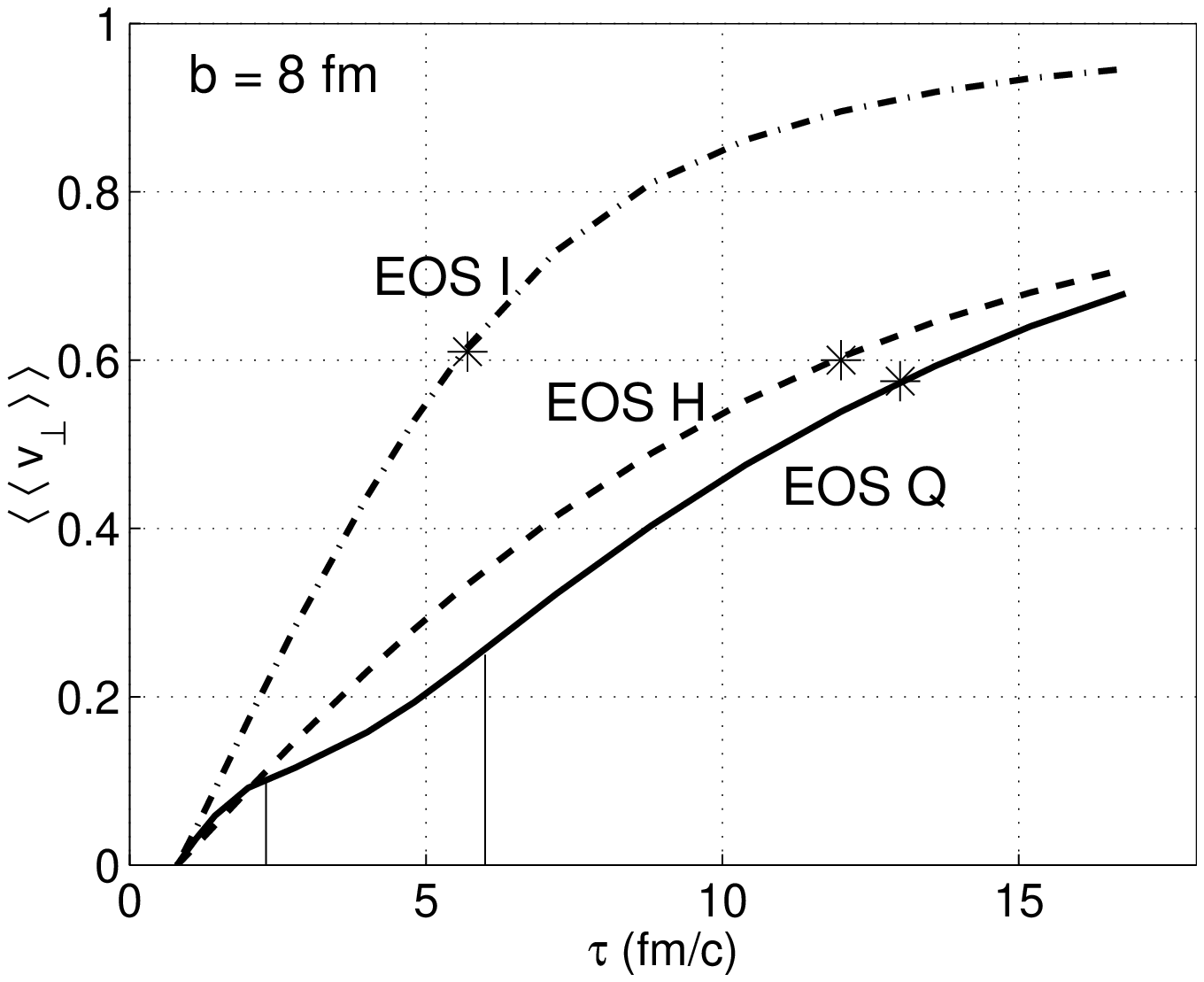}
Figure 1. Time evolution of the spatial eccentricity $\epsilon_x$
(top), the momentum anisotropy $\epsilon_p$ (middle), and the radial 
flow $\langle\!\langle v_{\perp} \rangle\!\rangle$ (bottom).
\label{epsvfig}
%\vspace*{2mm}
\end{minipage}
\end{minipage}
\hfill
%%%%%%%%%%%%%%%% Text to right of Fig. 1 %%%%%%%%%%%%%%%%%%%%%%%%%%%%%%%%%
\parbox[c]{9cm}{\vspace {1mm}
collisions. For each EOS these parameters and the decoupling 
temperature are fixed by a fit \cite{Kolb99} to the negative hadron and 
net proton $m_t$-spectra at midrapidity from central Pb + Pb collisions 
at 158 $A$GeV \cite{Appe99}. The spectra for non-central collisions
are then predicted without extra parameters. 

The hydrodynamic evolution provides the time-dependence of the 
matter in coordinate and momentum space. In non-central collisions,
the initial spatial deformation of the reaction zone in the transverse 
plane, characterized by its spatial eccentricity 
$
\epsilon_{x}{=}\frac{\langle\!\langle y^2{-}x^2\rangle\!\rangle}
                  {\langle\!\langle y^2{+}x^2\rangle\!\rangle}
$, leads to anisotropic pressure gradients and a preferred buildup
of transverse flow in the shorter $x$-direction \cite{Olli92}. 
($x$ lies inside, $y$ points orthogonal to the collision plane. 
$\langle\!\langle\dots\rangle\!\rangle$ denotes the energy density 
weighted spatial average at fixed time.) This leads to a growing 
flow anisotropy, characterized by 
$ 
\epsilon_{p}{=}\frac{\langle\!\langle T^{xx}{-}T^{yy}\rangle\!\rangle}
                  {\langle\!\langle T^{xx}{+}T^{yy}\rangle\!\rangle}
$. 
At freeze-out this hydrodynamic quantity is directly related to the 
elliptic flow $v_2=\langle \cos (2\varphi) \rangle$, defined by an 
average over the final particle momentum spectrum; for pions 
$\epsilon_p{\approx}2\,v_2$ \cite{Kolb99_2}. In contrast to 
$v_2$, $\epsilon_p$ can be studied as a function of time and 
gives access to the buildup of elliptic flow.

The developing stronger flow into the collision plane leads to 
a decrease of $\epsilon_x$ with time; the buildup of elliptic flow
thus slows down and eventually {\em shuts itself off}. This is clearly 
seen in the upper two panels of Fig.~1: $\epsilon_p$ saturates 
when $\epsilon_x$ passes through zero. For a hard EOS this happens 
faster than for a soft one; also, the total amount of elliptic flow 
which can be generated by a given EOS increases with its hardness 
$c_s^2 = {\partial e\over \partial p}$.}
%%%%%%%%%%%%%%%%%%%%% back to normal %%%%%%%%%%%%%%%%%%%%%%%%%%%%%%%%%%%%

Fig.~1 was computed for 158 $A$ GeV Pb+Pb collisions at $b=8$ fm.
For EOS~Q the vertical lines indicate when the center of the reaction 
zone goes from plasma to mixed phase and from mixed to hadron phase, 
respectively. One sees that 1/6 of the final elliptic flow is generated 
before the pure QGP phase disappears, 1/2 in the mixed phase, and about 
1/3 in the hadronic phase. The stars indicate the freeze-out point 
($T_{\rm dec}=120$ MeV). For EOS~I the system freezes out before the 
elliptic flow is fully developed; for EOS~H and EOS~Q the opposite is 
true. The radial flow, characterized by
$
  \langle\!\langle v_{\perp} \rangle\!\rangle{=}\big\langle\!\big\langle 
  \gamma \sqrt{v_x^2+v_y^2}\, \big\rangle\!\big\rangle\big/
  \langle\!\langle \gamma \rangle\!\rangle
$ (where $\gamma$ is the Lorentz factor), does not saturate: Fig.~1
shows that it continues to grow monotonously until freeze-out, even 
after the elliptic flow has saturated, due to the continued 
presence of {\em essentially azimuthally symmetric} radial pressure 
gradients.

Note the important role of the EOS: a softer EOS, especially its 
softening near a phase transition, delays the buildup of both radial 
and elliptic flow. It also reduces the maximally achievable value 
of the latter. At low energies, freeze-out (driven by cooling and 
expansion due to radial flow) happens before the elliptic flow has 
fully developed. To achieve fully developed elliptic flow lower
beam energies are required for softer equations of state 
\cite{Kolb99_2,Kolb99}. This reflects both the lower saturation 
value of $\epsilon_p$ for the softer EOS and the slower buildup 
of radial flow, resulting in more available time until freeze-out.

%%%%%%%%%%%%%%%%%%%%%%%%%%%%%%%%%%%%%%%%%%%%%%%%%%%%%%%%%%%%%%%%%%%%%%%%
\section{Transverse mass spectra and applicability of hydrodynamics}
%%%%%%%%%%%%%%%%%%%%%%%%%%%%%%%%%%%%%%%%%%%%%%%%%%%%%%%%%%%%%%%%%%%%%%%%

Collective flow affects the measurable momentum spectra of the 
produced particles. We showed in \cite{Kolb99_2} that our model is 
able to describe the measured asymmetries of the particle spectra;
the calculated elliptic flow $v_2{=}\langle \cos (2\varphi) \rangle$ 
at midrapidity agrees well with the published data \cite{Appe98}.
%%%%%%%%%%%%%%%%%% Figs. 2 and 3 %%%%%%%%%%%%%%%%%%%%%%%%%%%%%%%%%%%%%%
\vspace*{-1.1cm}
\begin{figure}[ht]
\begin{center}
\begin{minipage}[t]{7.8cm}
  \epsfxsize 7.60cm
  \epsfbox{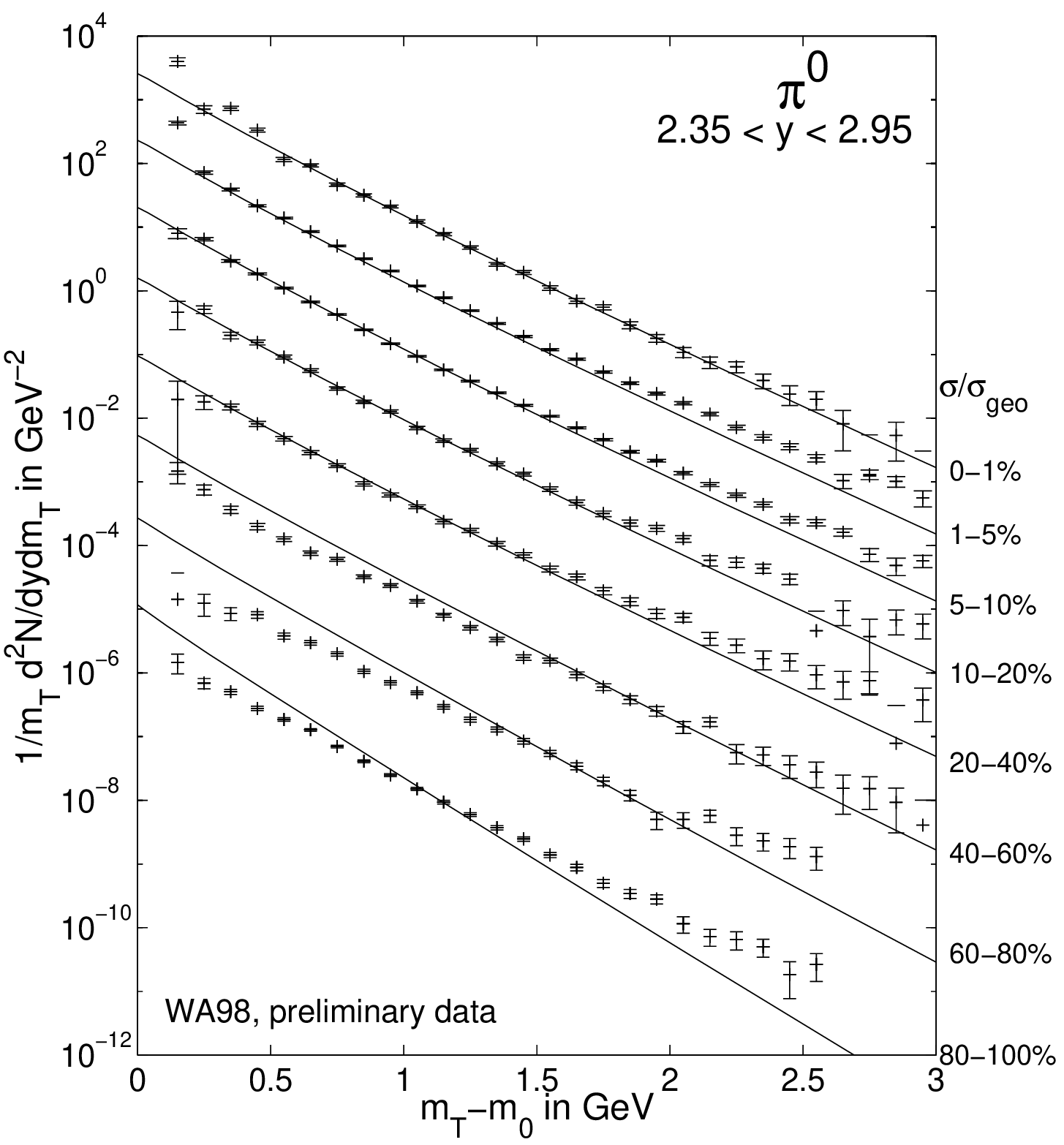}
Figure 2. Preliminary $\pi^0$ transverse mass spectra from Pb+Pb
collisions of varying centrality measured by WA98 \cite{Agga98}.
The lines show our hydro results.
\end{minipage}
\hfill
\begin{minipage}[t]{7.8cm}
  \epsfxsize 7.45cm
  \epsfbox{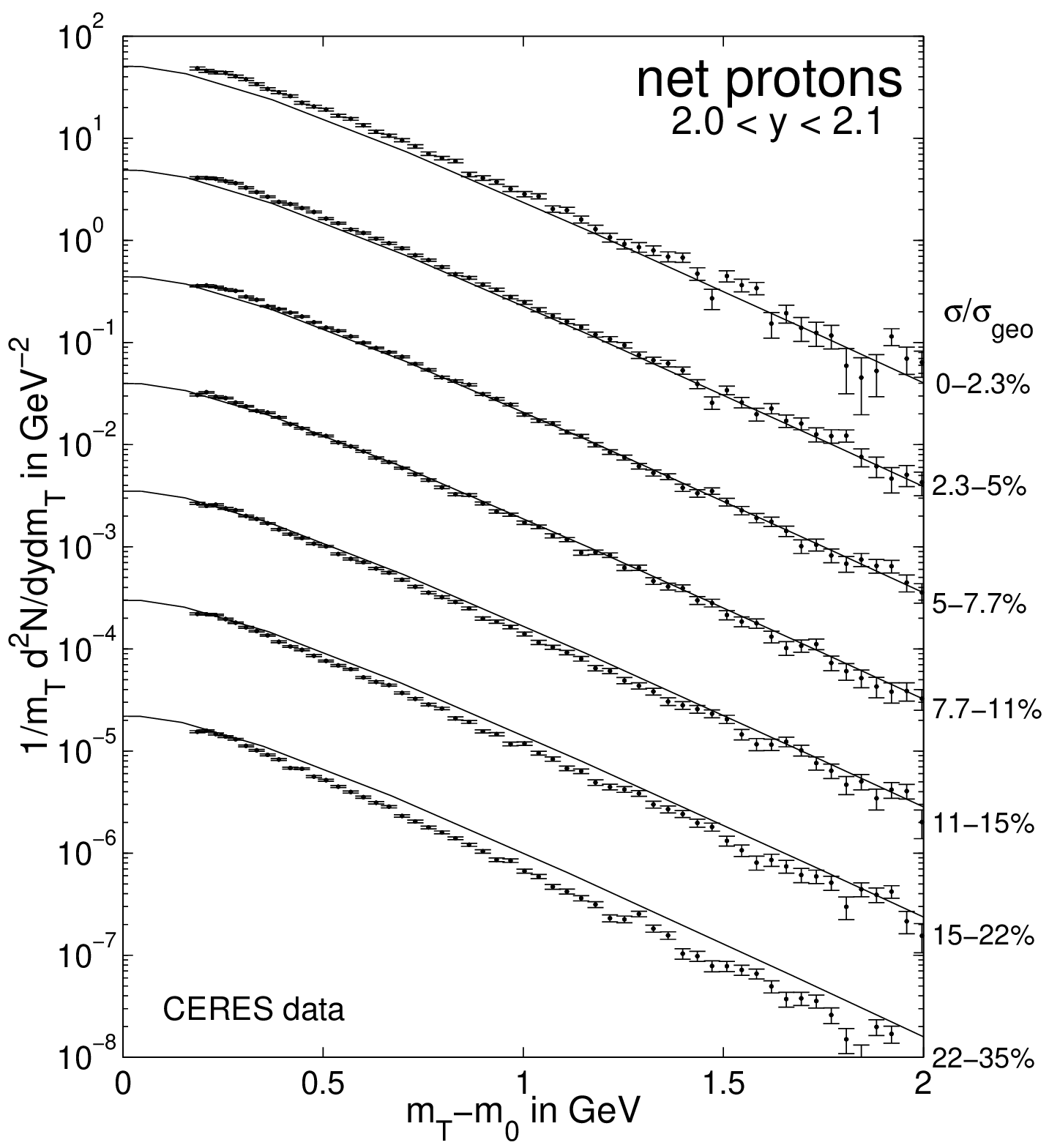}
Figure 3. Same as Fig.~2, but for net protons measured by the 
CERES Collaboration \cite{Cere98}. Note the different impact 
parameter binning in Figs.~2 and 3.
\end{minipage}
\end{center}
\end{figure}
\vspace*{-1cm}
%%%%%%%%%%%%%%%%%%%%%%%%%%%%%%%%%%%%%%%%%%%%%%%%%%%%%%%%%%%%%%%%%%%%%%%%
Here we discuss the impact parameter dependence of the azimuthally 
integrated transverse mass spectra and show that, once tuned to 
central collision data, hydrodynamics successfully reproduces the
magnitude and shape of the spectra up to impact parameters of about 
8-10 fm. Figs.~2 and 3 show preliminary data on transverse mass spectra 
of neutral pions (WA98 Collaboration \cite{Agga98}) and net protons
(CERES Collaboration \cite{Cere98}) from 158 $A$ GeV Pb+Pb collisions
of varying centrality. The lines indicate our hydrodynamical results
at midrapidity, obtained with EOS~Q and initial conditions tuned
to central collisions as described above. The WA98 data extend to very
peripheral collisions: the lowest spectrum in Fig.~2 corresponds to 
$b$=13 fm where hydrodynamics certainly looses its applicability. At 
such large impact parameters one does not observe the collision of two 
nuclei, but rather two dilute nucleon clouds penetrating each other. 
At smaller impact parameters the model fails in the high-$m_t$ region;
here hard scattering processes begin to dominate which cannot be 
modeled hydrodynamically. Up to $b\approx 10$ fm (bin 6 represents 
impact parameters up to 11 fm \cite{Agga98}) and transverse masses 
of about 2 GeV, however, hydrodynamics works very well, both for the
pion and net proton spectra. (For the latter the CERES data in Fig.~3 
do not extend to very peripheral collisions, the largest measured impact 
parameters corresponding to about 8.4 fm \cite{Cere98}.)

%%%%%%%%%%%%%%%%%%%%%%%%%%%%%%%%%%%%%%%%%%%%%%%%%%%%%%%%%%%%%%%%%%%%%%%%
\section{Summary}
%%%%%%%%%%%%%%%%%%%%%%%%%%%%%%%%%%%%%%%%%%%%%%%%%%%%%%%%%%%%%%%%%%%%%%%%

We have demonstrated the interplay between spatial eccentricity as
the driving force for generating momentum space asymmetries and
the back-reaction of the latter on the former. Comparison with 
measured spectra from Pb+Pb collisions with varying impact parameter 
showed that the hydrodynamical model successfully reproduces the 
data up to $b$=8-10 fm. The good quantitative agreement between data 
and model suggests rather rapid thermalization in the reaction zone. 
If final data confirm that the elliptic flow is essentially saturated 
in Pb+Pb collisions at the SPS, this would provide strong evidence for 
very early pressure in the system. In our calculations a large fraction 
of the finally observed elliptic flow is generated while the energy 
density exceeds the critical value $e_{\rm cr} = 1$ GeV/fm$^3$ for 
deconfinement. This confirms the suggestion \cite{Sorg99} that 
elliptic flow is a probe for the {\em early collision stage}. \\

We thank Th.~Peitzmann (WA98) and F.~Ceretto (CERES) for sending us 
their preliminary data prior to publication. P.K. wishes to express his 
gratitude to the CERN Summer Student Programme and thanks T.~Peeter's 
group for their warm hospitality.

%%%%%%%%%%%%%%%%%%%%% References %%%%%%%%%%%%%%%%%%%%%%%%%%%%%%%%%%%%%%%%

\end{document}